\documentclass[10pt,twocolumn,preprintnumbers,amsmath,amssymb,nofootinbib
,superscriptaddress]{revtex4-1}
\usepackage{graphicx,longtable,mathrsfs,color,array}
\usepackage[hidelinks]{hyperref}
\usepackage[usenames,dvipsnames]{xcolor} 
\usepackage{amssymb,amsmath,mathtools,mathrsfs,slashed} 
\usepackage{epsfig,subfigure,placeins,float} 
\usepackage{booktabs,longtable,ctable,multirow} 
\usepackage{exscale,relsize} 
\usepackage[normalem]{ulem} 
\usepackage{enumerate}
\usepackage{times, mathptmx} 
\usepackage[utf8]{inputenc}
\usepackage{color}
\usepackage{hyperref}
\usepackage{graphicx}
\usepackage{color}
\usepackage{comment}
\usepackage{graphicx,graphics}
\usepackage{bm}
\allowdisplaybreaks[1]

\begin{document}
\title{The speed of gravitational waves and black hole hair}
 \author{Oliver J. Tattersall}
\email{oliver.tattersall@physics.ox.ac.uk}
\affiliation{Astrophysics, University of Oxford, DWB, Keble Road, Oxford OX1 3RH, UK}
\author{Pedro G. Ferreira}
\email{p.ferreira1@physics.ox.ac.uk}
\affiliation{Astrophysics, University of Oxford, DWB, Keble Road, Oxford OX1 3RH, UK}
\author{Macarena Lagos}
\email{mlagos@kicp.uchicago.edu}
\affiliation{Kavli Institute for Cosmological Physics, The University of Chicago, Chicago, IL 60637, USA}
\date{Received \today; published -- 00, 0000}

\begin{abstract}
 The recent detection of GRB 170817A and GW170817 constrains the speed of gravity waves $c_T$ to be that of light, which severely restricts the landscape of modified gravity theories that impact the cosmological evolution of the universe. In this work, we investigate the presence of black hole hair in the remaining viable cosmological theories of modified gravity that respect the constraint $c_T=1$. We focus mainly on scalar-tensor theories of gravity, analyzing static, asymptotically flat black holes in Horndeski, Beyond Horndeski, Einstein-Scalar-Gauss-Bonnet, and Chern-Simons theories. We find that in all of the cases considered here, theories that are cosmologically relevant and respect $c_T=1$ do not allow for hair, or have negligible hair. We further comment on vector-tensor theories including Einstein Yang-Mills, Einstein-Aether, and Generalised Proca theories, as well as bimetric theories. 
 \end{abstract}

  \date{\today}
  \maketitle
\section{Introduction}\label{Sec:Introduction}
General Relativity (GR) is widely accepted to be the correct description of gravity at Solar System scales. In this regime, not only do its predictions show remarkable agreement with astrophysical data, but precise measurements of phenomena such as light deflection around the Sun, perihelion shift of Mercury, and others, rule out many modifications to GR. Nonetheless, GR exhibits weaknesses both at very high and very low energy regimes. 

At high energies, unavoidable singularities arise during gravitational collapses and the so-called renormalization problem limits the analysis of quantum states. At low energies, in particular, on cosmological scales, GR relies on the presence of an unknown component in order to explain the observed accelerated expansion of the Universe. The previous limitations suggest that GR may need modifications for both extreme energy regimes. Furthermore, modifications in the two regimes may be related as high energy corrections to GR might leak down to cosmological scales, showing up as low energy corrections. In this paper we explore possible connections between these two regimes. In particular, we will show how recent local constraints on the speed of gravity waves limit their solution for compact objects.     

In the past decades a large variety of gravity theories have been proposed \cite{Clifton:2011jh}. They have been extensively studied and constrained with cosmological data from CMB and large scale structure. However, recently strong constraints have been imposed with the detection of gravitational waves emission GW170817 from a neutron star binary merger by LIGO and VIRGO \cite{PhysRevLett.119.161101}, and its optical counterpart (the gamma ray burst GRB 170817A) \cite{2041-8205-848-2-L12,2041-8205-848-2-L13,2041-8205-848-2-L14,2041-8205-848-2-L15,2017Sci...358.1556C}. The delay of the optical signal was of 1.7 seconds, which places a stringent constraint on the propagation speed of gravity waves $c_T$. Indeed, it was found that $|c_T^2-1|<1\times 10^{-15}$ (with unity speed of light). As a result, a large class of modified gravity theories was highly disfavored as argued in \cite{Lombriser:2015sxa,Lombriser:2016yzn,Baker:2017hug,Creminelli:2017sry,Sakstein:2017xjx,Ezquiaga:2017ekz}. 

On the other hand, there have also been efforts to test gravity theories from observations of black holes \cite{Berti:2015itd, Cardoso:2016ryw, Johannsen:2015mdd}. GR predicts that black holes are solely characterized by a few ``charges" -- their mass $m$, angular momentum $a$ and electric charge $Q$ -- through what is known as the no-hair theorem. In extensions to GR additional degrees of freedom are usually introduced, which may add a new kind of charge to the spacetime solution. In this case, the metric carries additional information besides the mass, angular momentum and electric charge, and we say we are in the presence of a hairy black hole. These extra degrees of freedom will be carriers of fifth forces which may be detected from star orbits around the black holes \cite{Will:2007pp, Psaltis:2015uza, Hees:2017aal} or from the overall structure of accreting gas near the event horizon \cite{Psaltis:2010ca, Loeb:2013lfa, Johannsen:2016vqy}. These two examples correspond to non-dynamical tests which probe the stationary spacetime of a black hole. We note however, that even in the case where theories can avoid hair and have the same stationary black hole solutions as GR, new signatures may arise in dynamical situations; for example, when black holes are formed in a binary merger, the ringdown signal may carry a new set of modes that can be traced back to the new degrees of freedom \cite{Barausse:2008xv,Tattersall:2017erk, Berti:2018vdi}. In general, dynamical tests allow us to distinguish models that have the same stationary spacetime.  

In this paper we consider gravity models that have $c_T=1$ on a cosmological background -- where the extra degrees of freedom can play a significant role on cosmological scales -- and, as a first approach, expose the relation between the presence of hairy static black hole solutions and the speed of gravity waves. Before we proceed, it is important to clarify what type of ``hair" we will be considering here. We will consider hair to be any modification to GR that can be measured with non-dynamical tests. In particular, hair will be a permanent charge in a static, spherically symmetric, and asymptotically flat spacetime. When the metric profile is characterized by a new global charge (different to mass, spin or electric charge) we will say we have ``primary" hair, and if the profile has modifications that depend on the same charges as in GR, we say we have ``secondary" hair \cite{Herdeiro:2015waa, Bambi:2017}. This distinction is important to understand the number of independent parameters that fully determine the black hole solution, but both have physical consequences as they induce a geometry different to GR. 

We mention that in some cases it may be possible to construct ``stealth" black holes, where the geometry of spacetime is unchanged from the corresponding GR solution (i.e.~no new charge), but the black hole is dressed with a non-trivial additional field profile. For example, some stealth black hole solutions in scalar-tensor and vector-tensor theories can be found in {\cite{Babichev:2017guv,Heisenberg:2017hwb}. In this situation, non-dynamical tests will not be able to distinguish these models from GR and, therefore, for the purpose of this paper, these examples will not be considered as hairy solutions. 

With a clear definition of hair, we explore static, asymptotically flat black hole solutions on different modified gravity theories that are cosmologically relevant. We find that all scalar-tensor theories considered here have no hair at all or hair with negligible effects (although exceptions can be found in theories with no cosmological effects). Other models such as vector-tensor or bimetric theories more easily lead to hairy black holes regardless of their cosmological solution. We find examples with primary and secondary hair. Our study allows us to identify the theories which may lead to observable signatures on {\it both} cosmological and astrophysical scales, and can be used to build a roadmap for a coordinated study with future large scale structure surveys and gravitational wave observations.

No-hair theorems for GR and modified gravity (e.g.~\cite{Sotiriou:2015pka}) have been constructed under quite strict conditions, e.g.~the spacetime must be asymptotically flat and hair must be permanent. It is straightforward to break these conditions in a reasonable way \cite{Cardoso:2016ryw} and thus obtain hairy solutions, even in GR. For instance, changing the boundary conditions may lead to the Schwarzschild-De-Sitter solution which has a metric such that $g_{00}=1-2m/r+\Lambda r^3/3$, i.e.~an extra term proportional to $\Lambda$ which might be considered hair. Alternatively, in scalar-tensor theories, a time-dependent boundary condition for the scalar field can anchor hair on the surface of the black hole \cite{Jacobson:1999vr}. In the same way, more complex extra fields can be arranged to form hair. A notable example arises with a complex scalar field \cite{Herdeiro:2015waa} or with coupled dilaton-Maxwell systems \cite{PhysRevD.97.064032}. More recently, it has been shown that it is possible to construct solutions in which massive scalar fields hover around black holes for an extended period of time \cite{Cardoso:2011xi} leading to long-lived but not permanent hair. Given that we live in a cosmological spacetime with an abundance of fields, all of these examples show that hair can be easily present in black holes under reasonable assumptions. 

Nevertheless, all the mechanisms that have been proposed so far lead to very mild hair which, arguably, may be unobservable. For example, ``De Sitter" hair is remarkably weak compared to the usual Newtonian potential and any cosmological boundary condition that might lead to scalar hair will be highly suppressed. So if one can show that a theory must satisfy the no-hair theorem, it is extremely likely that any attempts at breaking it solely through changing the boundary condition will lead to effects which are too weak to be detected as a fifth force (although they might emerge in a stronger gravitational regime, like a black hole merger). This means that no-hair theorems are a useful guide to undertake a rough census of where to look in the panorama of gravitational theories.

This paper is organized as follows. In Section \ref{Sec:TensorSpeed} we discuss the speed of gravitational waves in the context of modified gravity. In Section \ref{Sec:ST} we focus on scalar-tensor theories with $c_T=1$ and discuss the presence of hairy static black hole solutions. In Sections \ref{Sec:Other} we discuss mainly vector-tensor theories as well as other gravitational models with $c_T=1$ that do evidence hair, such as bimetric theories. Finally, in Section \ref{Sec:conclusions} we summarise our results and discuss their consequences. 

Throughout this paper we will use natural units in which $G_N=c=1$.

\section{The speed of gravitational waves.}\label{Sec:TensorSpeed}
GR is a single metric theory for a massless spin-2 particle, and hence it propagates two physical degrees of freedom corresponding to two polarizations. On any given background spacetime, GR predicts that both polarizations propagate locally along null geodesics, and thus gravitational waves travel at the same speed as that of electromagnetic waves. This feature is particular of GR where Lorentz invariance is locally recovered, and hence all massless waves are expected to propagate at the same speed. However, such a feature can easily change in a theory of gravity where additional degrees of freedom are coupled to the metric in a non-trivial way. These additional fields can take special configurations in different backgrounds, and define a preferred direction that will spontaneously break local Lorentz invariance. In this case, there will be an effective medium for propagation of gravity waves, and their speed will change. Furthermore, depending on the configurations of the additional degrees of freedom, the speed of gravity waves could be anisotropic and even polarization dependent \cite{Tso:2016mvv}. 

The speed of gravity waves can be used to discriminate and test various modified gravity theories. This has been a topic of special interest in cosmology where a number of models have been proposed. In this case, the metric background is given by:
\begin{equation}
\bar{g}_{\mu\nu}=-dt^2+a(t)^2d\vec{x}^2,
\end{equation}
where $a(t)$ is the scale factor describing the expansion of the universe. Gravitational waves are described by small perturbations of the metric and thus we can write the total metric as:
\begin{equation}
g_{\mu\nu}=\bar{g}_{\mu\nu}+h_{\mu\nu},
\end{equation}
where $h_{\mu\nu}$ describes the amplitude of the waves and carries the information of all the metric polarizations. In this background, additional gravitational fields such as scalars or vectors will generically have a time-dependent solution which, even in a local frame, will define a preferred direction of time. It has been shown that for single-metric gravity models propagating a massless spin-2 particle, the action for gravity waves can generically be written as:
\begin{equation}\label{ActionTensors}
S=\frac{1}{2}\int d^3x dt \, M_*^2(t)\left[ \dot{h}_A^2  - c_T^2(t)( \vec{\nabla} h_A)^2\right],
\end{equation}
where we have expanded $h_{\mu\nu}$ in two polarization components $h_A$ with $A=+,\times$\footnote{Bimetric gravity theories will propagate additional tensor modes that will generically be coupled to $h$, and hence the action for gravity waves will be different to that in eq.~(\ref{ActionTensors}). Nevertheless, a similar analysis can be done to find $c_T$ (or some dispersion relation) in FRW.}. Here, $M_*$ is an effective Planck mass and $c_T$ is the propagation speed of gravity waves. Both of these quantities may in general depend on time, and thus in this case the speed will always be isotropic and polarization independent. It is usual that $c_T$ depends on the background solution of the additional gravitational degrees of freedom. 

Let us consider one particular example of a shift-symmetric quartic Horndeski gravity theory \cite{Horndeski:1974wa,Deffayet:2009wt} given by:
\begin{align}
S=&\frac{1}{2}\int d^4x \, \sqrt{-g}\left[  G(X) R \nonumber \right. \\
&\left. + G_{,X}(X) \left( (\Box \phi)^2-\nabla_\mu\nabla_\nu\phi \nabla^\mu\nabla^\nu\phi    \right) \right],
\end{align}
where $\phi$ is an additional gravitational scalar field, and $G$ is an arbitrary function of the kinetic term $X=-\frac{1}{2}\nabla_\mu \phi \nabla^\mu\phi$ and $G_{,X}$ its derivative with respect to $X$. On a cosmological background, we find $c_T^2=1/(1-\frac{2XG_{,X}}{G})$. One can see that, Taylor expanding, $G\simeq G_0+XG_{,X}$, if $XG_{,X}\ll G_0$ then, $c_T\simeq1$. This can occur if $XG_X$ is small but also if $G_0$ is large, i.e~if the contribution of the scalar field to the overall cosmological dynamics is negligible. In this paper we will consider the case when the contribution to the background dynamics is {\it not} negligible, i.e.~the extra degree of freedom has a relevant impact on cosmological scales.
 
There are, of course, cases in which the additional degrees of freedom do not affect the propagation speed of gravity waves. A particular, well-studied, example is Jordan-Brans-Dicke theory \cite{Jordan:1959eg,Brans:1961sx} given by:
\begin{equation} \label{JBDAction}
S=\frac{1}{2}\int d^4x \sqrt{-g}\left[  \phi R-\frac{\omega}{\phi} (\nabla_\mu \phi \nabla^\mu\phi) \right],
\end{equation}
where $\omega$ is an arbitrary constant. In this case $c_T=1$.  


\section{Scalar-Tensor Theories}\label{Sec:ST}
Scalar-tensor theories have been extensively studied in both the strong gravity and cosmological regime. Much effort in recent years has been put into researching general theories of a scalar field non-minimally coupled with a metric; from Horndeski gravity \cite{1974IJTP...10..363H}, to Beyond Horndeski \cite{Gleyzes:2014dya,Zumalacarregui:2013pma}, and Degenerate Higher Order Scalar Tensor (DHOST) theories \cite{Langlois:2017mdk}. Furthermore scalar-tensor theories are ubiquitous in that they appear as some limit of other theories of gravity, such as the decoupling limit of massive gravity \cite{deRham:2011by}. The well-posedness and hyperbolicity of scalar-tensor theories has been studied in \cite{Papallo:2017qvl,Papallo:2017ddx}.

We will focus on Horndeski and Beyond Horndeski theories in this section (as well as Chern-Simons \cite{Alexander:2009tp} and Einstein-Scalar-Gauss-Bonnet gravity \cite{Sotiriou:2014pfa}) and show the solutions of static black holes when $c_T=1$. In this paper, we will ignore DHOST theories due to the relative infancy of research into their black hole solutions. Cosmological consequences of the detection of GW/GRB170817 to DHOST theories has, however, been investigated in \cite{Langlois:2017dyl, Crisostomi:2017lbg}.

\subsection{Horndeski}\label{sechorn}
The most general action for scalar-tensor gravity with 2$^{nd}$ order-derivative equations of motion is given by the Horndeski action \cite{Horndeski:1974wa}:
\begin{align}
S=\int d^4x\sqrt{-g}\sum_{n=2}^5L_n,
\end{align}
where the Horndeski Lagrangians are given by:
\begin{align}
L_2&=G_2(\phi,X)\\
L_3&=-G_3(\phi,X)\Box \phi\\
L_4&=G_4(\phi,X)R+G_{4,X}(\phi,X)((\Box\phi)^2-\phi^{\mu\nu}\phi_{\mu\nu} )\\
L_5&=G_5(\phi,X)G_{\mu\nu}\phi^{\mu\nu}-\frac{1}{6}G_{5,X}(\phi,X)((\Box\phi)^3 \nonumber\\
& -3\box\phi \phi^{\mu\nu}\phi_{\mu\nu} +2 \phi_{\mu\nu}\phi^{\mu\sigma}\phi^{\nu}_{\sigma}),
\end{align}
where $\phi$ is the scalar field with kinetic term $X=-\phi_\mu\phi^\mu/2$, $\phi_\mu=\nabla_\mu\phi$, $\phi_{\mu\nu}=\nabla_\nu\nabla_\mu\phi$, and $G_{\mu\nu}=R_{\mu\nu}-\frac{1}{2}R\,g_{\mu\nu}$ is the Einstein tensor. The $G_i$ denote arbitrary functions of $\phi$ and $X$, with derivatives $G_{i,X}$ with respect to $X$.

For theories where the scalar field plays some role on the cosmological scales, the constraint $c_T=1$ imposes $G_{4,X}=0$ and $G_5$ has to be constant (in which case $L_5$ vanishes through Bianchi identity). Therefore, the resulting constrained Horndeski action is given by:
\begin{align}
S=\int d^4x \sqrt{-g}\left[G_4(\phi)R+G_2(\phi,X)-G_3(\phi,X)\Box\phi\right].\label{Sreduced}
\end{align}
In this case, we expect the cosmological energy density fraction of the scalar field to be $\Omega_\phi\sim O(1)$ due to our considering only \textit{cosmologically relevant} theories, where $\Omega_\phi$ is given for the above action by \cite{Bellini:2014fua}:
\begin{align}
\Omega_\phi=\frac{-G_2+2X\left(G_{2,X}-G_{3,\phi}\right)+6\dot{\phi}H_0\left(XG_{3,X}-G_{4,\phi}\right)}{6G_4H_0^2},
\end{align}
where $H_0$ is the value of the Hubble parameter today, $G_{i,\phi}$ denote derivatives of the functions $G_i$ with respect to $\phi$, and overdots denote derivatives with respect to time. 

We now proceed to analyse the black hole solutions arising from the action in equation (\ref{Sreduced}). Even though there is no no-hair theorem for generic $G_i$ functions, there are a number of theorems for restricted cases. We mention three distinct families of models. 

First, through a conformal transformation, the action in equation (\ref{Sreduced}) can be re-expressed in the form of GR plus a minimally coupled scalar field (with modified $G_2$ and $G_3$ \cite{Bettoni:2013diz}). The Lagrangian would then resemble a Kinetic Gravity Braiding model \cite{Deffayet:2010qz}. For $G_2=\omega(\phi)X,\, G_3=0$, the reduced Horndeski action can be seen to be in the form of generalised Brans-Dicke theories \cite{PhysRev.124.925}, for which a no hair theorem exists \cite{Sotiriou:2015pka}. 
Second, a no-hair theorem for K-essence (i.e.~$G_3=0$, $G_4=1$) is given in \cite{2014PhRvD..89h4056G}, provided that $G_{2X}$ and $\phi G_{2,\phi}$ are of opposite and definite signs.
Third, we mention that for shift-symmetric theories, which are invariant under $\phi \rightarrow \phi+$constant and hence $G_{i,\phi}=0$ in eq. (\ref{Sreduced}), the action takes the form of a minimally coupled scalar field with potentially unusual kinetic terms arising from $G_2(X)$ and $G_3(X)$. For such shift-symmetric theories, no-hair theorems exist for static black holes \cite{Hui:2012qt,Sotiriou:2015pka}. The outline of these no-hair theorems for shift-symmetric theories is given by:
(i) spacetime is spherically symmetric and static, and the scalar field shares the same symmetries; (ii) spacetime is asymptotically flat, $\frac{d\phi}{dr}\rightarrow 0$ when $r\rightarrow \infty$, and the norm of the Noether current associated to the shift symmetry is regular on the horizon; (iii) there is a canonical kinetic term $X$ in the action and the $G_{i,X}$ functions contain only positive or zero powers of $X$.

We note that this no-hair theorem is valid for all shift-symmetric Horndeski actions that satisfy the above conditions, even those that do not satisfy the constraint $c_T=1$. Focusing on a spherically symmetric and static spacetime, we can still have hairy black hole solutions by breaking assumptions (ii) or (iii) of this no-hair theorem. It is indeed expected that realistic situations of dark-energy models will break assumption (ii) as the scalar field is responsible for the late-time accelerated expansion of the universe or, more generally, the scalar field can lead to large-scale effects and thus $\frac{d\phi}{dr}$ does not necessarily vanish when $r\rightarrow \infty$. Examples like this can easily be realized by adding a time-dependent boundary condition to the scalar field \cite{Jacobson:1999vr} associated to the cosmological expansion. However, such a scalar hair would be highly suppressed and negligible in the vicinity of a black hole.  

We explore further cases that violate assumption (iii). A number of Lagrangians that break this assumption are discussed in \cite{Babichev:2017guv} but we will focus on the only two examples which still obey the constraint $c_T=1$ for Horndeski gravity. In addition, we will discuss a class of theories that even though they explicitly depend on $\phi$, are related to shift-symmetric theories via conformal transformations and therefore they also satisfy the no-hair theorem previously mentioned. 

\subsubsection{Quadratic term}

The first potentially hair-inducing term posited in \cite{Babichev:2017guv} is the addition of a square-root quadratic term to the canonical kinetic term, $G_2(X)=X+2\beta\sqrt{-X},\, G_4=\frac{1}{2}M_P^2$, where $\beta$ is an arbitrary constant. As we can see, in this case $G_{2,X}$ does contain negative powers of $X$ and thus hairy black holes may appear.  

First, we require the scalar field to be cosmologically relevant, and thus
\begin{align}\label{QuadraticOmega}
\Omega_\phi=\frac{-X}{3M_P^2H_0^2}\sim O(1).
\end{align}
Now, assuming a spherically symmetric and static ansatz for the metric and scalar field:
\begin{align}
ds^2=-h(r)dt^2+\frac{1}{f(r)}dr^2+r^2d\Omega^2,\; \phi=\phi(r)\label{ansatz}
\end{align}
and requiring that the radial component of the Noether current $J^r=\frac{1}{\sqrt{-g}}\frac{\delta S}{\delta\left(\partial_r \phi\right)}=0$ (to ensure a regular current on the horizon, as required in assumption 2 above) we find:
\begin{align}
-X&=\frac{1}{2}f(r)\left(\frac{d\phi}{dr}\right)^2=\beta^2.
\end{align}
We can then solve the field equations (provided in \cite{Babichev:2017guv}) for the metric function $f(r)$ to find:
\begin{align}
f(r)=h(r)=1-\frac{2m}{r}+\frac{\beta^2}{3M_P^2}r^2.
\end{align}
Thus, we find a stealth Schwarzschild-AdS black hole of mass $m$ with an effective negative cosmological constant $\Lambda_{\text{eff}}=-{\beta^2}/{M_P^2}$ (assuming real $\beta$ and hence $\beta^2>0$.). If we required asymptotic flatness then this model would have the same solution as GR, and there would be no hair. Relaxing that assumption, we note that cosmologically relevant scalar fields satisfy eq.~(\ref{QuadraticOmega}) and we then expect $\Lambda_{\text{eff}}\sim H_0^2$. Therefore, hair would be negligible near the black hole.

\subsubsection{Cubic term}

A second possibility analysed in \cite{Babichev:2017guv} is the introduction of a logarithmic cubic term with $G_2=X,\,G_3=\alpha M_P\log(-X),\, G_4=M_P^2/2$, where $\alpha$ is an arbitrary dimensionless constant. Again, we see that in this example, $G_{3,X}$ has negative powers of $X$. If the scalar field is to have cosmological relevance, we need:
\begin{align}
\Omega_\phi=\frac{X+6\dot{\phi}H_0\alpha M_P}{3M_P^2H_0^2}\sim O(1).
\end{align}
Again, using the ansatz given by eq.~(\ref{ansatz}) we find the following expression for ${d\phi}/{dr}$ by requiring $J^r=0$:
\begin{align}
\frac{d\phi}{dr}=&\;-\alpha M_P\left(\frac{1}{h(r)}\frac{dh(r)}{dr}+\frac{4}{r}\right),
\end{align}
To proceed we assume that $h(r)=f(r)$. Making use of the field equations calculated in \cite{Heisenberg:2017hwb} (translating from a vector-tensor theory to a shift-symmetric scalar-tensor theory such that $A_\mu\rightarrow\partial_\mu\phi$, i.e.~$A_0=X_0=0,\, A_1={d\phi}/{dr}$), we find the following hairy solution:
\begin{align}
ds^2=&-\left(1-\frac{2m}{r}+\frac{c}{r^{4+\frac{1}{\alpha^2}}}\right)dt^2+\left(1-\frac{2m}{r}+\frac{c}{r^{4+\frac{1}{\alpha^2}}}\right)^{-1}dr^2\nonumber\\
&+\frac{r^2}{1+4\alpha^2}d\Omega^2,
\end{align}
where we have rescaled $r$ and $t$ by constant pre-factors to obtain a Schwarzschild-like solution.
If we imposed asymptotic flatness, we would find that the solution cannot have hair as the metric line element does not approach Minkowski when $r\rightarrow \infty$ due to the factor of ${1}/({1+4\alpha^2})$ in the angular part. Thus, we cannot construct a static, spherically symmetric, asymptotically flat solution with scalar hair in this theory (under the assumption that $h(r)=f(r)$). Furthermore, the fact that $G_3$ generically diverges for $X=0$ is suggestive that Minkowski space is not a solution for this theory, and therefore this model does not seem to be viable.

\subsubsection{Conformally shift-symmetric theories}

We now proceed to discuss models that depend explicitly on $\phi$, and hence break the assumptions of the shift-symmetric no-hair theorem. While such models could generically lead to hairy black holes, here we analyse a special class that is conformally related to shift-symmetric theories, and thus avoids scalar hair. 

In the prototypical scalar-tensor theory of gravity, Brans-Dicke theory, it is well known that the theory can be recast from the `Jordan frame' (in which a non-minimal coupling between the scalar and curvature exists) into that of GR with a minimally coupled scalar field through the use of a conformal transformation \cite{Bettoni:2013diz,Sotiriou:2015pka}. The trade off is that, in this so-called `Einstein frame' where the non-minimal coupling between the scalar field and curvature has been eliminated, any additional matter fields no longer couple solely to the metric but also to the gravitational scalar field. However, if we work in a vacuum then both Jordan and Einstein frames are entirely physically equivalent. We can use this same analysis to show that theories in the Jordan frame of the type:
\begin{align}\label{SConformalShift}
S_J=\frac{M_P^2}{2}\int d^4x\;\sqrt{-g}\;&\left[\phi R +\phi^2F_2(X/\phi)-\phi F_3(X/\phi)\Box\phi\right.\nonumber\\
&\left.-V(\phi)\right]
\end{align}
can be transformed from the Jordan frame into the Einstein frame through the conformal transformation $\tilde{g}_{\mu\nu}=\phi g_{\mu\nu}$. Here, $F_i$ are arbitrary functions of $X/\phi$ and $V$ is a potential for the scalar field. Eq.~ (\ref{SConformalShift}) leads to the following action in the Einstein frame:
\begin{align}
S_E=\frac{M_P^2}{2}\int d^4x\;\sqrt{-\tilde{g}}\;&\left[ \tilde{R} +F_2(\tilde{X})+2\tilde{X}F_3(\tilde{X})-F_3(\tilde{X})\tilde{\Box}\phi\right.\nonumber\\
&\left.-\phi^{-2}V(\phi)\right].
\end{align}
In the case of vanishing potential $V=0$, the Einstein frame action $S_E$ is clearly shift symmetric in the scalar field $\phi$. Thus, via the no-hair theorem in \cite{Hui:2012qt}, static black hole solutions for $\tilde{g}_{\mu\nu}$ should be the same as in GR with no scalar hair. As a consequence, solutions for $g_{\mu\nu}$ from $S_J$ will not have hair either. 

For cosmologically relevant models, $\phi$ will have a fractional energy density given by:
\begin{widetext}
\begin{align}
\Omega_\phi=\frac{V-\phi^2 F_2 +2X\left(\phi^2 F_{2,X}-F_3-\phi F_{2,\phi}\right)+3\dot{\phi} H_0\left(2\phi X F_{3,X}-M_P^2\right)}{3M_P^2H_0^2\phi}\sim O(1).
\end{align}
\end{widetext}

Generalizing to the case with $V\neq0$, it is known that if $F_3=0$ we then have a K-essence model and static black hole solutions will not have hair provided that $V_{,\phi\phi}>0$ and $F_{2,X}>0$ (these conditions can be interpreted as constraining the scalar field to be stable and to satisfy the null energy condition \cite{2014PhRvD..89h4056G,Sotiriou:2015pka}). 

For non-zero $V$ and $F_3$, the no-hair condition can be shown to be (through integrating $(\phi^{-2}V)_\phi \mathcal{E}_\phi$ from the horizon to spatial infinity, with $\mathcal{E}_\phi$ being the equation of motion for the scalar field \textit{in the Einstein frame} \cite{Maselli:2015yva}):
\begin{widetext}
\begin{align}
&(F_2+2\tilde{X}F_3)_{,\tilde{X}}+f(r)\frac{d\phi}{dr} F_{3\tilde{X}}\left(\frac{1}{2h(r)}\frac{dh(r)}{dr}-\frac{2}{r}\right)\geq0\;\text{and}\;(\phi^{-2}V)_{\phi\phi}\geq 0\nonumber\\
\text{or}\; & (F_2+2\tilde{X}F_3)_{,\tilde{X}}+f(r)\frac{d\phi}{dr} F_{3\tilde{X}}\left(\frac{1}{2h(r)}\frac{dh(r)}{dr}-\frac{2}{r}\right)\leq0\;\text{and}\;(\phi^{-2}V)_{\phi\phi}\leq 0,
\end{align}
\end{widetext}
where $h(r),\,f(r)$ are the metric functions in the spherically symmetric ansatz given by eq.~(\ref{ansatz}).

\subsection{Beyond Horndeski}

Horndeski gravity can be extended by the addition of terms which lead to higher order derivative equations of motion, but without an extra propagating degree of freedom \cite{Gleyzes:2014dya,Zumalacarregui:2013pma}. The beyond Horndeski terms are given by 
\begin{align}
L_4^{BH}&=F_4(\phi,X)\epsilon^{\mu\nu\rho}_{\sigma}\epsilon^{\mu'\nu'\rho'\sigma}\phi_\mu\phi_{\mu'}\phi_{\nu\nu'}\phi_{\rho\rho'}\\
L_5^{BH}&=F_5(\phi,X)\epsilon^{\mu\nu\rho\sigma}\epsilon^{\mu'\nu'\rho'\sigma'}\phi_\mu\phi_{\mu'}\phi_{\nu\nu'}\phi_{\rho\rho'}\phi_{\sigma\sigma'}
\end{align}
The condition $c_T=1$ generalizes to:
\begin{equation}
F_5=0, \quad G_{5,X}=0, \quad G_{4,X}-G_{5,\phi}=2XF_4.
\end{equation}
Note that by setting $F_4=0$ in the above equation (i.e. recovering Horndeski theory), the condition for $c_T=1$ appears to be $G_{4,X}=G_{5,\phi}$ rather than $G_{4,X}=G_{5,\phi}=0$ as stated in section \ref{sechorn}. This is not inconsistent, as Horndeski theories with $G_{4,X}=G_{5,\phi}$ will indeed result in $c_T=1$ \cite{Baker:2017hug,Creminelli:2017sry,Sakstein:2017xjx,Ezquiaga:2017ekz}. As discussed in \cite{Baker:2017hug}, however, in the Horndeski case we require both $G_{4,X}$ and $G_{5,\phi}$ to vanish \textit{independently} rather than relying on any finely tuned cancellation between the two terms. On the other hand, for Beyond Horndeski theories we can make use of the presence of the additional free function $F_4$ to cancel the contributions of $G_{4,X}$ and $G_{5,\phi}$ in $c_T$, thus preserving a richer landscape of viable theories with $c_T=1$.

Similarly to the Horndeski case, we first require the scalar energy density parameter to be cosmologically relevant, i.e.~$\Omega_\phi\sim O(1)$, with $\Omega_\phi$ given by: 
\begin{widetext}
\begin{align}
\Omega_\phi=\frac{-G_2+2X\left(G_{2,X}-G_{3,\phi}\right)+6H_0\dot{\phi}\left(XG_{3,X}-G_{4\phi}-2XG_{4,\phi X}\right)+24H_0^2X^2\left(F_4+G_{4,XX}\right)-48H_0^2 X^2\left(2F_4+XF_{4,X}\right)}{6H_0^2\left(G_4-2XG_{4,X}+XG_{5\phi}\right)}
\end{align}
\end{widetext}

In \cite{Lehebel:2017fag}, it is shown that shift-symmetric Horndeski \textit{and} Beyond Horndeski have no hair for a regular, asymptotically flat spacetime, with canonical kinetic term $X$ in action and positive powers of $G_{i,X}$ and $F_{i,X}$. In what follows, we focus again on models that break the last assumption. We investigate two terms given in \cite{Babichev:2017guv} that respect the constraint $c_T=1$. 

\subsubsection{Square root Quartic models}

We first consider including a $\sqrt{-X}$ term in $G_4$ (with the choice of $F_4$ corresponding to the above conditions that lead to $c_T=1$):
\begin{align}
G_2=X,\;G_4=\frac{1}{2}M_P^2+\gamma\sqrt{-X},\;F_4=\frac{\gamma}{4(-X)^{\frac{3}{2}}},
\end{align}
where $\gamma$ is an arbitrary constant.

For this choice of $G_i, F_i$, the condition for the scalar field to be cosmologically relevant is given by:
\begin{align}
\Omega_\phi=\frac{X-6H_0^2\gamma\sqrt{-X}}{3H_0^2M_P^2}\sim O(1).
\end{align}
Assuming a spherically symmetric ansatz for the metric as in eq.~(\ref{ansatz}), we find two branches of solutions for $X$:
\begin{eqnarray}
X(r)=&\;0 \quad \Rightarrow\quad  \frac{d\phi}{dr}=0,
\end{eqnarray}
or
\begin{eqnarray}
 X(r)=& -\left(\frac{4\gamma^2+M_P^2r^2}{3\gamma r^2}\right)^2
\Rightarrow  \frac{d\phi}{dr}= \sqrt{\frac{2}{f(r)}}\frac{4\gamma^2+M_P^2r^2}{3\gamma r^2},
\end{eqnarray}
respectively. We see that the first branch results in a solution with a constant scalar field, i.e.~no-scalar hair, resulting in regular GR black holes. We thus try to find solutions for the metric functions $f(r)$ and $h(r)$ for the second branch of solutions for $X(r)$. We find the following for $f(r)$:
\begin{align}
f(r)=\frac{64 \gamma ^6+9 \gamma ^2 c_1 r^3-M_P^6 r^6+45 \gamma ^2 M_P^4 r^4+144 \gamma ^4 M_P^2
   r^2}{9 \left(\gamma  M_P^2 r^2-8 \gamma ^3\right)^2}.
\end{align}
For large $r$, $f(r)\sim r^2$, this solution is clearly not asymptotically flat. Thus we have not been able to construct an asymptotically flat spherically symmetric black hole solution with scalar hair for this model. 
 
\subsubsection{Purely Quartic models}
Purely quartic models are proposed in \cite{Babichev:2017guv} (i.e.~only $G_4$ and $F_4$ non-zero and with no canonical kinetic term for the scalar field). One such model that obeys the $c_T=1$ constraint is given by:
\begin{align}
G_4=&\frac{1}{2}M_P^2+\sum_{n\geq2}2a_n\frac{(X-X_0)^{n+1}}{(n+1)(n+2)}\left[(n+1)X+X_0\right]\\
F_4=&\sum_{n\geq2}a_n(X-X_0)^n,
\end{align}
where $X=X_0$ is the constant value of the background scalar field kinetic term around the black hole:
\begin{align}
\frac{d\phi}{dr}=\pm \sqrt{-\frac{2X_0}{f(r)}},
\end{align}
thus leading to non-trivial scalar field profile for $X_0\neq0$. 

The cosmological density parameter for the scalar field in this case is:
\begin{widetext}
\begin{align}
\Omega_\phi=\frac{-8X^2\sum_{n\geq2}a_n\left(X-X_0\right)^n}{M_P^2+4\sum_{n\geq2}\frac{a_n\left(X-X_0\right)^n}{n+2}\left[(3-2n)X^2-\frac{1}{n+1}X_0\left(nX+X_0\right)\right]}
\end{align}
\end{widetext}
and we expect it to be of order 1. 

The black hole solution of this model is a stealth black hole, where the spacetime geometry is given by the appropriate GR solution, but the scalar field takes a non-trivial profile (as shown above). For a stealth Schwarzschild black hole, the scalar field is thus given by \cite{Babichev:2017guv}:
\begin{align}\label{STstealth}
\phi(r)=\sqrt{-2X_0}\left[\sqrt{r^2-2mr}+m\log\left(r-m+\sqrt{r^2-2mr}\right)\right].
\end{align}
The above profile for the scalar field is regular everywhere outside the horizon of the black hole, with $\phi \sim r$ as $r\rightarrow \infty$.

Since the spacetime geometry is the same as that of GR, we do not expect to be able to distinguish this model through non-dynamical tests such as analyses on orbits of stars or electromagnetic imaging of the accretion flow around the black hole. Nevertheless, we do expect to see a difference in dynamical situations. In particular, it has been suggested that while scalar-tensor theories with $\phi=$constant may not lead to any new signature during the inspiral of two black holes, the no-hair theorem can be pierced if the scalar field has some dynamics \cite{Healy:2011ef}. In the case of stealth black holes, the scalar field will have a non-trivial initial profile as in eq.~(\ref{STstealth}) which may trigger a subsequent dynamical evolution which may lead to dipole gravitational wave radiation which in turn will change the evolution of the emitted GW waveform phasing compared to that of GR.

\subsection{Einstein-Scalar-Gauss-Bonnet}

In four dimensions, the Gauss-Bonnet (GB) term is a topological invariant, and as such the addition of the GB term to the usual Einstein-Hilbert action of GR does not affect the equations of motion. If, however, the GB term is non-minimally coupled to a dynamical scalar field in the action, the dynamics are significantly altered. Consider the action of Einstein-Scalar-Gauss-Bonnet (ESGB) gravity \cite{Sotiriou:2014pfa}:
\begin{align}
S_{GB}=\int d^4x\sqrt{-g}&\left[\frac{M_P^2}{2}R-\frac{1}{2}g^{\mu\nu}\phi_\mu\phi_\nu-V(\phi)\right.\nonumber\\
&\left.-\frac{1}{2}\xi(\phi)R^2_{GB}\right],\label{SGB}
\end{align}
where we have introduced a scalar field $\phi$ with potential $V(\phi)$, which for simplicity we will neglect from now on, and a coupling function $\xi(\phi)$. In addition, we have defined $R_{GB}^2=R_{\mu\nu\alpha\beta}R^{\mu\nu\alpha\beta}-4R_{\mu\nu}R^{\mu\nu}+R^2$ as the Gauss-Bonnet term. 

It is well known that models given by eq.~(\ref{SGB}) can produce scalar hair on black hole backgrounds in both the static \cite{Sotiriou:2014pfa,2017CQGra..34f4001B,Antoniou:2017hxj} and slowly rotating \cite{Pani:2009wy,Maselli:2014fca,Maselli:2015tta,Kleihaus:2014lba,Kleihaus:2011tg,Ayzenberg:2014aka} regimes. However, on a cosmological background they lead to a modified speed of gravity waves. Indeed, we find that 
\begin{align}
c_T=1+4\frac{\left(H\dot{\phi}-\ddot{\phi}\right)\xi^{\prime}-\dot{\phi}^2\xi^{\prime\prime} }{M_P^2-4H\dot{\phi}\xi^{\prime}}, \label{CTGB}
\end{align}
where a prime denotes a derivative with respect to $\phi$ (see also \cite{Gong:2017kim}). Equivalently, eq.~(\ref{SGB}) can be recast into the form of Horndeski gravity with the following choice of functions $G_i$:
\begin{align}
G_2=&X-V+4\xi^{\prime\prime\prime\prime}X^2\left(\log{X}-3\right)\nonumber\\
G_3=&2\xi^{\prime\prime\prime}X\left(3\log{X}-7\right)\nonumber\\
G_4=&\frac{M_P^2}{2}+2\xi^{\prime\prime}X\left(\log{X}-2\right)\nonumber\\
G_5=&2\xi^\prime\log{X}.
\end{align}
In this form, it is clear the ESGB gravity does not conform to the constraints of \cite{Baker:2017hug,Creminelli:2017sry,Sakstein:2017xjx,Ezquiaga:2017ekz} that $G_{4,X}=0=G_5$ to ensure that $c_T=1$. Furthermore, it is well known that the case of coupling to the GB term is a loophole in the no-hair theorems for shift-symmetric Horndeski theories \cite{Sotiriou:2014pfa,Sotiriou:2015pka,Maselli:2015yva}.

Assuming that the scalar field is cosmologically relevant, we find that $\Omega_\phi\sim O(1)$ where 
\begin{widetext}
	\begin{align}
	\Omega_\phi=\frac{V+X+4\left(\xi^{\prime\prime\prime\prime}-\xi^{\prime\prime\prime}\right)X^2\left(3-\log{X}\right)+24H_0X\left(H_0\xi^{\prime\prime}+3\dot{\phi}\xi^{\prime\prime\prime}\right)}{3H_0^2\left(M_P^2+4\xi^{\prime\prime}X\log{X}-4\xi^{\prime}H_0\dot{\phi}\right)}.
	\end{align}
\end{widetext}
We then impose $c_T=1$ for a generic background evolution. Therefore, from eq.~(\ref{CTGB}) we get $\xi^{\prime}=\xi^{\prime\prime}=0$, in which case the GB term decouples from the scalar field $\phi$ and we obtain GR with a minimally coupled scalar field plus a GB term. In this case, as mentioned above, the addition of the GB term to the usual Einstein-Hilbert term represents nothing more than the addition of a total divergence that leaves the equations of motion unaffected. Thus the constraint on $c_T$ rules out the possibility of ESGB gravity having any cosmological relevance. As discussed above, the scalar field $\phi$ could avoid modifying $c_T$ at an observable level only if it is assumed to be incredibly sub-dominant on cosmological scales, i.e.~if $\Omega_\phi\ll 1$.

As a counter-example to the above discussion of the cosmological impact of ESGB gravity, the following theory is studied in \cite{Granda:2018tzi}:
\begin{align}
S=\int d^4x\sqrt{-g}&\left[\frac{M_P^2}{2}R-\frac{1}{2}g^{\mu\nu}\phi_\mu\phi_\nu-V(\phi)\right.\nonumber\\
&\left.+F_1(\phi)G^{\mu\nu}\phi_\mu\phi_\nu-\frac{1}{2}F_2(\phi)R^2_{GB}\right],
\end{align}
with string-inspired exponential forms for $F_1, F_2$, and $V$. It is shown in \cite{Granda:2018tzi} that this theory, including \textit{both} scalar coupling to the GB term and scalar derivative coupling to the Einstein tensor, can admit de-Sitter like and power-law cosmological solutions whilst maintaining $c_T=1$. This theory is clearly not shift-symmetric and so the no-hair theorem of \cite{Hui:2012qt} is not applicable, whilst a coupling to the GB term is known to produce black holes with scalar hair \cite{Sotiriou:2014pfa,2017CQGra..34f4001B,Antoniou:2017hxj}, or at least that are unstable to spontaneous scalarization \cite{Doneva:2017bvd,Silva:2017uqg}. Thus we expect that, in general, black holes in this string-inspired theory can possess scalar hair and satisfy current constraints on the speed of gravity waves, although with no cosmological effects.

\subsection{Chern-Simons}
Chern-Simons (CS) gravity \cite{Alexander:2009tp} is characterised by the addition of the Pontryagin invariant, $^*RR=\frac{1}{2}\epsilon^{\mu\nu\alpha\beta}R^\lambda_{\sigma\alpha\beta}R^\sigma_{\lambda\mu\nu}$ to the standard Einstein-Hilbert term of the action. The Pontryagin invariant can be coupled to either a dynamical or non-dynamical scalar field, leading to two different formulations of the theory. For concreteness, consider the dynamical formulation of CS gravity
\begin{align}
S_{CS}=\int d^4x \sqrt{-g}\left[\frac{M_P^2}{2}R-\frac{1}{2}g^{\mu\nu}\phi_\mu\phi_\nu+\alpha f(\phi)^*RR\right],
\end{align}
where $\alpha$ is a constant coupling parameter and $f(\phi)$ is an arbitrary function of the scalar field.

In \cite{Yagi:2016jml,2010PhRvD..82d1501G,2005PhRvD..71f3526A} it is shown that in CS gravity, gravitational waves propagate at the speed of light on conformally flat background spacetimes such as FRW. As such, \cite{Yagi:2016jml} postulates that it is not possible to constrain CS gravity purely through the propagation speed of gravitational waves. Regardless, static and spherically symmetric black holes in CS gravity do not have hair and admit the same solutions as in GR \cite{Molina:2010fb}.

\section{Other theories}\label{Sec:Other}
We now consider other theories which go beyond the broad span of scalar-tensor theories. As one expects, the moment one considers fields with more ``structure" (i.e.~more indices), there is a greater possibility of non-trivial coupling with the metric which, in turn, can lead to black hole hair. 

\subsection{Einstein gravity with Maxwell, Yang-Mills and Skyrme fields}
The simplest theory with a vector field corresponds to an Einstein-Maxwell system. In this case, the black hole solution is Kerr-Newman which, besides mass and spin, is characterized by an electric charge. While this black hole solution is not considered hairy, we will start by mentioning this case (and its non-abelian extensions) to get a flavor of what to expect in the case of fields with more structure.

The Einstein-Maxwell theory is given by
\begin{eqnarray}
S=\int d^4x\sqrt{-g}\left[\frac{M^2_P}{2}R-\frac{1}{4}F_{\alpha\beta}F^{\alpha\beta} \right],
\end{eqnarray}
where $F_{\mu\nu}=\partial_\mu A_\nu - \partial_\nu A_\mu$ is the Maxwell tensor associated with a vector field $A^\alpha$. In this model, we have that the FRW cosmological fractional energy density of the vector field is:
\begin{eqnarray}\label{EMOmega}
\Omega_A=\frac{F_{0\alpha}F^{\alpha}_{\phantom{\alpha}0}+\frac{1}{4}F_{\alpha\beta}F^{\alpha\beta}}{3M^2_{\rm Pl}H_0^2}\sim\frac{|\vec{B}|^2}{H_0^2M^2_{\rm Pl}},
\end{eqnarray}
where ${\vec B}$ is the associated magnetic field. In this isotropic background, one has that $A^{\mu}=(A_0(t),{\vec 0})$ and gravity wave propagation will be direction independent with exactly $c_T=1$. Furthermore, in this case, the cosmological evolution is exactly the same as that of GR as the magnetic field vanishes ($\vec{B}=0$) in this homogeneous background, and hence $\Omega_A=0$ from eq.~(\ref{EMOmega}). 

If the vector field is cosmologically relevant, taking into account the fact that the electrical conductivity of the universe is large, we expect the presence of magnetic fields which will lead to anisotropies. On the one hand, in the case of stochastic magnetic fields, the metric may be locally anisotropic but too weak to affect local gravitational wave propagation. 

On the other hand, in the case of a global magnetic component, the vector field will have a spatial dependence ${\vec A}\neq0$ and the universe will be anisotropic. Therefore, the propagation of gravitational waves will be direction dependent \cite{Hu:1978td}. Current constraints on global anisotropy (and homogeneous magnetic fields) from the cosmic microwave background are remarkably tight \cite{Bunn:1996ut,Barrow:1997mj,Ade:2015bva} and we will enforce strict isotropy in what follows (although it is conceivable that multiple measurements of $c_T$ in different directions might improve these constraint). 

The black hole solutions for the Einstein-Maxwell system are Kerr-Newman, which are fully characterized by three charges -- mass $m$, spin $a$ and electric charge $Q$ -- and with a non-trivial profile for the vector potential $A^\alpha$. In the spherically symmetric case (where $a=0$), one has the Reissner-Nordstrom solution given by: 
\begin{eqnarray}
h(r)=1-\frac{2m}{r}+\frac{Q^2}{4\pi \epsilon_0 r^2} \ \mbox{and} \ \
A_0(r)=\frac{Q}{r},
\end{eqnarray}
where $\epsilon_0$ is the vacuum permittivity constant. Here, we have again used a metric of the form of eq.~(\ref{ansatz}) and $A^\mu=(A_0(r),A_1(r),0,0)$, with $A_1$ being an unphysical gauge mode. 

The above description can be extended to the case where the gauge field is non-abelian -- the Einstein-Yang-Mills system -- or the vector field has a stronger non-linear self coupling -- the Einstein-Skyrme system; in this case we are considering genuine hair. In both of these cases, new phenomena can come into play. While, on the whole, the energy density of the fields can remain sub-dominant at cosmological scales, non-perturbative structures (topological and non-topological defects) can in principle make a non-trivial contribution to the overall energy density and to the global isotropy of space (although, generally, these effects are expected to be weak). Again, we can enforce $c_T=1$ yet still allow for non-abelian hair. Notable examples can be found in the Einstein-Yang-Mills case \cite{Bartnik:1988am,Volkov:1989fi} which combine solitonic cores with long range forces; in the Einstein-Skyrme cases there is a range of solutions combined with solitonic states \cite{Shiiki:2005pb,Gudnason:2016kuu,Adam:2016vzf}. 

\subsection{Einstein-Aether}

Generalized Einstein-Aether is a gravity theory where the metric is coupled to a unit time-like vector field, dubbed the aether. This model provides a simple scenario for studying effects of local Lorentz symmetry violation. In particular, the vector field defines a preferred frame where boosts symmetry is broken but rotational symmetry is still preserved. 
The action describing this model is given by:
\begin{eqnarray}
S=\int d^4x\sqrt{-g}\left[\frac{M^2_P}{2}R+{\cal F}(K)+\lambda(A^\mu A_\mu+1) \right],
\end{eqnarray}
where $\lambda$ is a Lagrange multiplier that forces the vector field to be unit time-like. Also, ${\cal F}(K)$ is an arbitrary function of the kinetic term $K$ given by $K=c_1\nabla_\mu A_\nu \nabla^\mu A^\nu+c_2(\nabla_\mu A^\mu)^2+c_3\nabla_\mu A_\nu \nabla^\nu A^\mu$ (with $c_i$ constants) (an additional quartic term in $A^\mu$ contributing to $K$ is sometimes considered).

The cosmological consequences of Einstein-Aether have been extensively explored in \cite{Zlosnik:2006zu,Zuntz:2010jp,Audren:2014hza}. The case where the aether field can play the role of either dark matter or dark energy were explored in \cite{Zuntz:2010jp} where the existing cosmological constraints ruled it out as an alternative to a cosmological constant. In the case of the standard Einstein-Aether case (with ${\cal F}(K)\sim K$), current Solar System constraints \cite{Jacobson:2008aj} combined with binary pulsar constraints \cite{Yagi:2013qpa,Yagi:2013ava} place $|c_1|,|c_3|\le 10^{-2}$ and $c_2\le 1$. Cosmological constraints allow $\Omega_{A}\sim 0.3$ so the aether field can still have a non-negligible contribution to the cosmological evolution \cite{Zuntz:2008zz}.

The propagation speed of gravitational waves on a cosmological background is such that $c_T^2=1/[1+(c_1+c_3){\cal F}_{,K}]$, and thus the constraint on $c_T$ implies $c_1=-c_3$. This means that $K$ is reduced to a canonical kinetic term (an ``$F^2$" term) supplemented by a $(\nabla_\mu A^\mu)^2$ term.

For models which respect $c_T=1$, the condition that the aether field $A^\mu$ has cosmological relevance is thus given by:
\begin{align}
\Omega_A=\frac{-{\cal F}}{3M_P^2H_0^2(1-3c_2{\cal F}_{,K})}\sim O(1),
\end{align}
 where ${\cal F}_{,K}$ denotes the derivative ${\cal F}$ with respect to $K$.

Little has been done on black hole solutions for general ${\cal F}(K)$ and thus we will restrict ourselves to the standard Einstein-Aether case. Black hole solutions with hair have been found in such a case, that are regular, asymptotically flat and depend on only one free parameter \cite{Eling:2006ec,Eling:2006df,Konoplya:2006ar,Barausse:2011pu,Barausse:2013nwa}. For instance, in the case where $c_3=0$, $c_2$ must satisfy the condition:
\begin{equation}
c_2=-\frac{c_1^3}{2-4c_1+3c_1^2},
\end{equation}
where $c_1$ is the only free parameter of the black hole solution. We can use a spherically symmetric ansatz for the line element as in eq.~(\ref{ansatz}). The full solution must be found numerically, but an analytical perturbative solution can be given when $x=2m/r\ll 1$:
\begin{align}
h(r)&=1+x+(1+c_1/8)x^2+...\, ,\\
f(r)^{-1}&=1-x-c_1/48 x^2+...
\end{align}
We note that this is an example of primary hair, where the spacetime geometry is different to that of GR and, furthermore, the solution depends on an additional independent free parameter $c_1$. 

More general solutions satisfying $c_3=-c_1$ (and hence $c_T=1$) are expected to have the same behaviour \cite{Eling:2006ec}. This shows that BH solutions always have hair, regardless of additional cosmological constraints. While in static black holes deviations from GR are typically of a few percent (only exceeding $10\%$ for some region of the viable parameter space) \cite{Barausse:2013nwa}, generalizations to spinning black holes may offer better prospects for observing hair in these models.

\subsection{Generalised Proca}\label{secVT}
The generalised Proca theory \cite{Heisenberg:2014rta,Tasinato:2014eka,Allys:2015sht,Allys:2016jaq,Jimenez:2016isa} is given by
\begin{equation}
S=\int d^4x\; \sqrt{-g}\left\{ F+\sum_{i=2}^6\mathcal{L}_i[A_\alpha, g_{\mu\nu}]\right\}, 
\end{equation}
where $\mathcal{L}_i$ are gravitational vector-tensor Lagrangians given by:
\begin{widetext}
\begin{align}\label{VT}
&\mathcal{L}_2=G_2(X,F,Y),\nonumber\\
&\mathcal{L}_3=G_3(X)\nabla_{\mu}A^\mu,\nonumber\\
&\mathcal{L}_4=G_4(X)R+G_{4X}\left[ (\nabla_{\mu}A^\mu)^2 - \nabla_{\mu}A^\nu\nabla_{\nu}A^\mu \right],\nonumber\\
&\mathcal{L}_5=G_5(X)G_{\mu\nu}\nabla^{\mu}A^\mu-\frac{1}{6}G_{5X}\left[ (\nabla_{\mu}A^\mu)^3-3(\nabla_{\alpha}A^\alpha)(\nabla_{\mu}A^\nu\nabla_{\nu}A^\mu) \right.\nonumber \\
& \left. +2 \nabla_{\mu}A^\alpha \nabla_{\nu}A^\mu \nabla_{\alpha}A^\nu \right] - g_5(X)\tilde{F}^{\alpha\mu}\tilde{F}_{\beta\mu}\nabla_{\alpha}A^\beta,\nonumber\\
&\mathcal{L}_6=G_6(X)L^{\mu\nu\alpha\beta}\nabla_\mu A_\alpha \nabla_\alpha A_\beta +\frac{1}{2}G_{6X}\tilde{F}^{\alpha\beta}\tilde{F}^{\mu\nu}\nabla_{\alpha}A_\mu \nabla_{\beta}A_\nu, 
\end{align}
\end{widetext}
which are written in terms of 6 free functions $G_2$, $G_3$, $G_4$, $G_5$, $g_5$, and $G_6$. We can define the following tensors:
\begin{align}
& F_{\mu\nu}=\nabla_\mu A_\nu -\nabla_\nu A_\mu ,\\
& \tilde{F}^{\mu\nu}=\frac{1}{2}\epsilon^{\mu\nu\alpha\beta}F_{\alpha\beta},\\
& L^{\mu\nu\alpha\beta}=\frac{1}{4}\epsilon^{\mu\nu\rho\sigma} \epsilon^{\alpha\beta\gamma\delta}R_{\rho\sigma\gamma\delta},
\end{align}
where $R_{\rho\sigma\gamma\delta}$ is the Riemann tensor and $\epsilon^{\mu\nu\alpha\beta}$ is the Levi-Civita antisymmetric tensor. The 5 free parameters $G_i$ are free functions of the following scalar quantities of the previously defined tensors:
\begin{align}
& X = -\frac{1}{2}A_\mu A^\mu,\\
& F =- \frac{1}{4}F^{\mu\nu}F_{\mu\nu},\\
& Y = A^\mu A^\nu F_\mu{}^\alpha F_{\nu\alpha}.
\end{align}

The conditions to have $c_T=1$ are $G_{4,X}=G_{5,X}=0$ (with the term proportional to $G_{\mu\nu}\nabla^\mu A^\nu$ vanishing due to the Bianchi identity). In this case, the condition for cosmological relevance of the vector field $A^\mu$ is given by \cite{DeFelice:2016yws}:
\begin{align}
\Omega_A=\frac{-G_2+G_{2,X}A_0^2+3G_{3,X}H_0A_0^3}{6G_4H_0^2}\sim O(1),
\end{align}
where $A^\mu=\left(A_0(t),\vec{0}\right)$.

Most of the theories that satisfy $c_T=1$ are of the form of GR with a minimally coupled vector field possessing `exotic' kinetic terms, in which case hairy black holes are to be expected. It can be easily shown that spherically symmetric BHs can indeed have hair. For instance, \cite{Heisenberg:2017hwb} shows the solution when $G_3\not=0,\,G_4={M_P^2}/{2}$, in which case:
\begin{align}
f(r)=\left(1-\frac{m}{r}\right)^2,\; A_0=\sqrt{2}M_{Pl}\left(1-\frac{m}{r}\right),\;A_1=0,
\end{align}
where we have again assumed a metric ansatz given by eq.~(\ref{ansatz}) (with $h(r)=f(r)$), and that $A^\mu=\left(A_0(r),A_1(r),0,0\right)$. This resembles an extremal Reissner-Nordstrom black hole with a `charge' that depends on mass. This is an example of secondary hair, where the spacetime metric is different to that of GR, but both theories depend on the same number of independent free parameters. 
 
In contrast, since the Lagrangian $\mathcal{L}_6$ couples the vector field non-minimally to curvature through $L^{\mu\nu\alpha\beta}$ this Lagrangian corresponds to an intrinsic vector mode, and as such does not contribute to the background equations of motion for a homogenous and isotropic cosmological background (where $A^\mu=\left(A_0(t),\vec{0}\right)$) \cite{Nakamura:2017dnf}. In this case, with non-minimal coupling to curvature, one solution is that of a stealth Schwarzschild black hole with a non-trivial profile for the background vector field \cite{Heisenberg:2017hwb}:
\begin{align}
f(r)=1-\frac{2m}{r},\;A_0=\text{\it}{const},\;A_1=\frac{\sqrt{A_0^2-2X_0f(r)}}{f(r)},
\end{align}
where $X=X_0=\text{\it}{const}$.
As shown in \cite{Tattersall:2017erk}, the QNMs of this stealth Schwarzschild black hole will be unaffected from the usual GR spectrum due to $F_{\mu\nu}=0$ for the above vector profile.

\subsection{Scalar-Vector-Tensor}
We now consider theories in which there are two additional fields to the metric: a scalar and a vector. A specific model was analysed in \cite{Heisenberg:2018vti}, where a shift-symmetric scalar field $\phi$ \textit{and} a $U(1)$ gauge invariant vector field $A^\mu$ are coupled \cite{Heisenberg:2018acv}. Specifically, the action of interest is given by:
\begin{align}
S=\int &d^4x\,\sqrt{-g}\left[\frac{M_P^2}{2}R+X+F+\beta_3 \tilde{F}^\mu_{\;\rho} \tilde{F}^{\nu\rho}\phi_\mu \phi_\nu \right.\nonumber\\
&\left.+ \beta_4 X^{n-1} \left(XL^{\mu\nu\alpha\beta}F_{\mu\nu}F_{\alpha\beta}+\frac{n}{2}\tilde{F}^{\mu\nu}\tilde{F}^{\alpha\beta}\phi_{\mu\alpha}\phi_{\nu\beta}\right)\right],\label{SVT}
\end{align}
where $\beta_3$ and $\beta_4$ are arbitrary constants, $X=-\phi_\mu \phi^\mu/2$ is the scalar kinetic term as in scalar-tensor theories, and with all other terms being defined as in Section \ref{secVT}. 

Given that the vector field strength $F_{\mu\nu}$ vanishes on isotropic cosmological backgrounds (with $A^\mu=(A_0(t),\vec{0}))$, the action given by eq.~(\ref{SVT}) reduces to that of GR with a minimally coupled, massless scalar field in cosmological settings. The speed of gravitational waves $c_T$ will, therefore, be equal to unity in these theories, thus satisfying the constraint determined by GRB 170817A and GW170817. 

Black hole solutions in this model were studied in \cite{Heisenberg:2018vti}, where asymptotically flat, static and spherically symmetric black holes with hair are found for $\beta_4=0$ and for $\beta_4\neq0$ in the cases of $n=0$ or $1$. In all of the cases studied, modified Reissner-Nordstrom-like solutions with global charges $m$ and $Q$ are found, with the black hole further endowed with a secondary scalar hair sourced by the vector charge $Q$. 

\subsection{Bigravity}
We now consider bimetric theories. The only non-linear Lorentz invariant ghost-free model is given by the deRham-Gabadadze-Tolley (dRGT) \cite{deRham:2010kj, deRham:2010ik, Hassan:2011zd} massive (bi-)gravity action:
\begin{align}
S&=\; \frac{M_g^2}{2}\int d^4x\; \sqrt{-g}R_g +\frac{M_f^2}{2}\int d^4x\; \sqrt{-f}R_f \nonumber \\
&- m^2M_{g}^2\int d^4x\; \sqrt{-g}\sum_{n=0}^4\beta_n e_n\left(\sqrt{g^{-1}f}\right),
\end{align}
where we have two dynamical metrics $g_{\mu\nu}$ and $f_{\mu\nu}$ with their associated Ricci scalars $R_g$ and $R_f$, and constant mass scales $M_g$ and $M_f$, respectively. Here, $\beta_n$ are free dimensionless coefficients, while $m$ is an arbitrary constant mass scale. The interactions between the two metrics are defined in terms of the functions $e_n (\mathbb{X})$, which correspond to the elementary symmetric polynomials of the matrix $\mathbb{X}=\sqrt{g^{-1}f}$. 

This bigravity action generally propagates one massive and one massless graviton, with the field $g_{\mu\nu}$ being a combination of \textit{both} modes. There is one special case where $M_f/M_g\rightarrow \infty$, and only the massive graviton propagates (while the metric $f_{\mu\nu}$ becomes a frozen reference metric). As discussed in \cite{Baker:2017hug}, constraints on the speed of gravity waves lead to bounds on the graviton mass of the order $m \lesssim 10^{-22}$eV, which are weak compared Solar System fifth force constraints of order $m \lesssim 10^{-30}$eV. As long as $m\sim H_0$ we expect the massive graviton to have some cosmological relevance. 

Regarding black hole solutions, massive gravity (with one dynamical metric) has some static solutions, although they have been found to be problematic as they can describe infinitely strongly coupled regimes or have singular horizons. One well-behaved solution was proposed recently in \cite{Rosen:2017dvn}, for a time-dependent black hole.  

On the other hand, massive bigravity has a much more rich phenomenology with a number of possible stationary solutions (static, rotating, and with or without charge) (see a review in \cite{Babichev:2015xha}). Focusing on asymptotically flat solutions, it is possible to have Schwarzschild or Kerr solutions for both metrics \cite{Babichev:2014tfa}. However, these solutions are generically unstable, although they can still fit data as long as $m \lesssim 5\times 10^{-23}$eV \cite{Brito:2013wya}. Hairy static solutions can also be found for some parameter space \cite{Brito:2013xaa}. Using the ansatz in eq.~(\ref{ansatz}) for the metric $g_{\mu\nu}$ and the following ansatz for $f_{\mu\nu}$ :
\begin{equation}
ds^2_f=-P(r)dt^2+\frac{1}{B(r)}dr^2 +U(r)^2d\Omega^2. 
\end{equation}
Here, there are five independent functions $\{h, f, P, B, U\}$ to be determined by the equations of motion. However, due to the presence of a Bianchi constraint, there are only three independent functions $\{ f, B, U\}$ satisfying first-order differential equations. The complete solution must be found numerically, but an expansion can be made for $r\rightarrow \infty$ \cite{Brito:2013xaa}: 
\begin{align}
f(r)^{1/2}&=1-\frac{c_1}{2r}+\frac{c_2(1+r\mu)}{2r}e^{-r\mu},\nonumber \\
Y(r)&= 1-\frac{c_1}{2r}-\frac{c_2(1+r\mu)}{2r}e^{-r\mu} \nonumber, \\ 
U(r)& = r+ \frac{c_2(1+r\mu+r^2\mu^2)}{r^2\mu^2}e^{-r\mu}
\end{align}
where $c_i$ are integration constants, $\mu=m\sqrt{1+M_g^2/M_f^2}$, and $Y$ is a proxy function for $B$ given by $Y=U'/B^{1/2}$. While the constant $c_1$ may be identified with the mass of the black hole, $c_2$ is a new charge that adds a Yukawa-type suppression to the metric due to the massive graviton. 

Here we have mentioned some possible black hole solutions for bimetric theories, but we note that since these solutions are not unique, it is not clear what the physical spacetime and the outcome of gravitation collapse will be. Future simulations on non-linear gravitational collapse should allow us to find the physical solution.

\section{Conclusion}\label{Sec:conclusions}

Modifications to general relativity may affect the evolution of the universe and lead to cosmologically observable effects. The range of possible modifications has been drastically reduced with the discovery of GW170817 and the resulting constraint on the speed of gravitational waves. We have looked at the reduced space of theories to see which of them will lead to distinctive signatures around black holes, specifically black hole hair. By looking for observable signatures of that hair and combining them with constraints from current and future cosmological surveys, it should be possible to further narrow down the span of allowed modifications to general relativity and, if the data points that way, single out new physics.

We have focused on scalar-tensor theories. Of all theories, these are the most thoroughly understood and, furthermore, emerge as low energy limits of other, more intricate theories. Not only is there a reasonably general classification of scalar-tensor theories, but there is also a comprehensive body of work on black holes and black hole hair arising in them. As was shown in \cite{Baker:2017hug,Creminelli:2017sry,Sakstein:2017xjx,Ezquiaga:2017ekz}, the discovery of GW170817 places severe constraints on these models. We have found that, generally, and in the cases where they have been studied more carefully, these theories do not have hair for static, spherically symmetric, and asymptotically flat black holes. Specific examples that were constructed to have hair (as suggested in \cite{Babichev:2017guv}), in the case where they contribute cosmologically and satisfy $c_T=1$, do not have hair. We found that the case where Einstein-Scalar-Gauss-Bonnet gravity is cosmologically relevant, it is ruled out by the GW170817 constraint, while Chern-Simons gravity is left unconstrained (and furthermore, known to have hair in the slowly rotating regime \cite{Yunes:2009hc,Konno:2009kg,Yagi:2012ya,Konno:2014qua,Stein:2014xba,McNees:2015srl}). We also looked at other theories, primarily involving vectors, and found that in that case it is possible for them to satisfy the various cosmological conditions and still have black holes with hair for spherically symmetric and rotating black holes. We also discussed bimetric theories, which allow for hairy and non-hairy asymptotically flat black holes, and understanding which solution describes physical setups require further work on gravitational collapse.  

Our analysis is limited in scope in the sense that we have not considered the most general actions allowed. For example we have not considered combinations of the Beyond Horndeski models studied in \cite{Babichev:2017guv}. We have done this for two reasons. The first reason is that these models were constructed as proofs of concept without any strong underlying physical motivation -- questions of analyticity arise in the limit where $X\rightarrow 0$. The second reason is that the equations of motion become vastly more complicated with multiple non-analytic leading terms which means it is difficult to obtain solutions which can be easily interpreted and classified. Lacking more general results (such as the Galileon no-hair theorem of \cite{Hui:2012qt}) it is always plausible that theories, which satisfy the constraints we impose and lead to hair, exist.

Nevertheless, our analysis is useful for determining how to move forward with the theories we looked at. Given their cosmological relevance, we take for granted that they will be thoroughly tested when the next generation of cosmological data is made available. What we can now do is determine how to combine these cosmological tests with non-dynamical tests in the strong-field regime. In the cases where the black holes do have hair, one would look for evidence of a fifth force for example in the accreting material or nearby objects. 

For theories where black holes have no hair the situation is more complex. In that case, the only observations that might lead to data which allow us to constrain the extra fields are dynamical tests which include inspirals of binary black holes, extreme mass-ratio inspirals (EMRI) or ringdown of single black holes. In GR, due to the strong equivalence principle, the orbital motion and the gravitational wave signal of binary inspirals or EMRI only depend on the masses and spins of the objects involved. On the contrary, in modified gravity theories this principle is violated, and the additional degrees of freedom will determine the effective gravitational coupling constant which will depend on the new field or its derivatives at the location of the relevant object (e.g.~see \cite{Will:2004xi, Mirshekari:2013vb, Horbatsch:2011ye, Healy:2011ef} for binary inspirals in scalar-tensor theories and a general analysis in \cite{Yunes:2009ke,Berti:2005qd}, as well as \cite{Yunes:2011aa, Sopuerta:2009iy} for EMRI in scalar-tensor theories). In the case of ringdown, one would expect that the violent event would have excited any putative extra degree of freedom (such as a scalar or a vector). And while the end state might be a Kerr-Newman solution, the perturbations around this background (i.e.~the quasi-normal modes) should contain information about the extra degrees of freedom \cite{Tattersall:2017erk}. It would be interesting further work to study the specific quasi-normal-mode signatures of the theories investigated here that do not exhibit hairy black hole solutions (yet still abide by the $c_T$ constraint).

Finally, given that we have found that most scalar-tensor theories abide to the no-hair theorem and present trivial constant profiles around black holes, it would be interesting to explore black hole solutions in the presence of screening. It has been argued that models satisfying Solar System constraints (i.e.~which hide the presence of fifth forces in the weak-field limit) do so through screening. The main screening mechanisms which have been advocated are the Vainshtein \cite{Vainshtein:1972sx}, chameleon \cite{Khoury:2003rn} and symmetron \cite{Hinterbichler:2010es} mechanisms, all of which suppress the fifth force compared to the Newtonian force, depending on the local environment. The current approach is to assume that the self gravity of compact objects is sufficiently substantial that it decouples the scalar charge from the mass -- in the limit of a black hole, the scalar charge is set to zero \cite{Hui:2012jb}. However, little has been done to construct screened black hole solutions by, for example, violating the condition of asymptotic flatness. Such an analysis might give us insight on the presence of hair in more realistic setups. A particularly interesting scenario might arise in the case of a binary neutron star merger, such as GW170817. There, screened compact objects (neutron stars) end up forming a black hole; if indeed the black hole has no hair (and no screening) one might expect that the process of shedding the scalar field could lead to an observable effect. Alternatively, if the black hole adopts the screening mechanism, it would be useful to understand what is the final, stable, solution and how it jibes with the no-hair theorem. 

We have entered a new era in gravitational physics in which multiple regimes can be tested with high precision. While multi-messenger gravitational wave physics has grown to prominence, we believe it should now also include other, significantly different, arenas; from the cosmological, through the galactic all the way down to astrophysical and compact objects, a wide range of observations can be brought together to construct a highly precise understanding of gravity.

\section*{Acknowledgments}
\vspace{-0.2in}
\noindent We are grateful to E.~Bellini, V.~Cardoso, and N.~Sennett for useful discussions. OJT was supported by the Science and Technology Facilities Council (STFC) Project Reference 1804725. OJT also thanks the Max Planck Institute for Gravitational Physics (Albert Einstein Institute) for hosting him whilst part of this work was completed, funded by an STFC LTA grant. PGF acknowledges support from STFC, the Beecroft Trust and the European Research Council. ML is supported at the University of Chicago by the Kavli Institute for Cosmological Physics through an endowment from the Kavli Foundation and its founder Fred Kavli.
\bibliography{RefModifiedGravity}

\end{document}